\documentclass[twocolumn]{aastex63}
\usepackage[utf8]{inputenc}
\usepackage{amsmath, mathtools}
\usepackage{graphicx}
\usepackage{wrapfig}
\usepackage{natbib,color,ulem,hyperref}
\makeatletter
\renewcommand\paragraph{\@startsection{paragraph}{4}{\z@}%
            {-2.5ex\@plus -1ex \@minus -.25ex}%
            {1.25ex \@plus .25ex}%
            {\normalfont\normalsize}}
\makeatother
\def\g{\, \rm g}
\def\K{\, \rm K}

\def\s{\, \rm s}
\def\km{\, \rm km}

\def\m{\, \rm m}

\newcommand{\z}{\color{red}}
\newcommand{\q}{\color{red}}
\newcommand{\w}{\color{purple}}
\newcommand{\y}{\color{magenta}}
\newcommand{\x}{\sout}

\shorttitle{shadowed disk}
\shortauthors{Qian \& Wu}

\begin{document}

\title{Shadows Wreak Havocs in Transition Disks}

\author[0000-0002-6794-7480]{Yansong Qian}
\email{yansong.qian@mail.utoronto.ca}
\author[0000-0003-0511-0893]{Yanqin Wu}
\affiliation{Department of Astronomy \& Astrophysics, University of 
Toronto}

\begin{abstract}
We demonstrate that shadows cast on a proto-planetary disk can drive it eccentric. Stellar irradiation dominates heating across much of these disks, so an uneven illumination can have interesting dynamical effects. 
%the outer disk, even a 10\% temperature change can cause significant disruption.
Here, we focus on transition disks. We carry out  3D Athena++ simulations, using a constant thermal relaxation time to describe the disk's response to changing stellar illumination.  
We find that an asymmetric shadow, a feature commonly observed in real disks, perturbs the radial pressure gradient and distorts the fluid streamlines into a set of twisted ellipses. Interactions between these streamlines  have a range of consequences. For a narrow ring, an asymmetric shadow can sharply truncate its inner edge, possibly explaining the steep density drop-offs observed in some disks and obviating the need for massive perturbers. For a wide ring, such a shadow can dismantle it into two (or possibly more) eccentric rings. These rings continuously exert torque on each other and drive gas accretion at a healthy rate, even in the absence of disk viscosity. Signatures of such twisted eccentric rings may have already been observed as, e.g.,  twisted velocity maps inside gas cavities. We advocate for more targeted observations, and for a better understanding on the origin of such shadows.
%h a constant relaxation time to an azimuthally varying equilibrium temperature. 
% Meanwhile, the shadow also drives gas accretion in wide rings, reaching $\sim 10^{-8}\, M_\odot \mathrm{yr}^{-1}$ for a disk with standard mass. 
\end{abstract}

\section{Introduction}
\label{sec:intro}

%{\w make more explicit discussion on why $m=2$ doesn't drive anything for us; but surprisingly for others, cite the other two papers}

Over the past decade, high resolution observations have revealed a wide variety of sub-structures in proto-planetary disks, including cavities \citep{Long2018}, rings \citep{alma2015}, arcs \citep{marel2013} and spirals \citep{Kurtovic2018}. Such features are likely closely related to planet formation, but it remains unclear if planets are the cause or the effect.

A prominent type of sub-structures is an inner cavity. Disks with this morphology, known as transition disks,  can often be recognized from their spectral energy distribution \citep{Strom1989,Calvet2002},  
%{\y are these helpful references:'However, it is not clear clear that TDs are always old, nor that all old PPDs pass through the TD phase (Owen 2016; Manzo-Mart´ınez et al. 2020; van der Marel 2023, and references therein)'.} 
as the inner clearing leads to a  deficit in infrared radiation \citep[for reviews of transition disks, see][]{Espaillat2014,Owen2016,vandermarel2023}.
The origin of these cavities is currently unclear. While planetary perturbers are often invoked, 
simulations have found that even multiple planets have difficulties in opening wide and deep gaps \citep{Zhu2011}.  The discovery of transition disks in very young sources \citep{Sheehan2017} also places more strain on the planet hypothesis.
% {\w sheehan+2018: class 0/I/flat spectrum source already contains transitional disk}
% Sheehan & Eisner (2017) found the first case of an embedded “transition disk”, with a single large cavity in the center, while Sheehan (2018) found the first embedded disk analogous to HL Tau in GY 91, with three gaps within its large disk. Additional rings, gaps, and asymmetries have subsequently been found in a handful of additional disks (Ohashi et al., 2022, Yamato et al., 2023, Sheehan et al., 2020, 2022a), including in the disks of the well known protostars L1489 IRS and L1527 IRS.
%,Crida2007}.

In addition to an unclear origin, transition disks also present other puzzles.  For instance, they commonly exhibit sharp inner edges
%as revealed by CO and millimetre observations 
\citep[e.g.,][]{marel2015b,Dong2017}. In the ring of MASS J16042165–2130284 (shortened as J1604 in this work), densities of big dust, small dust and possibly gas drop by a few orders of magnitudes over a distance less than 10au, or a couple local scale heights  \citep{Dong2017}. %Other rings may show similar sharp drops if observed at higher resolution \citep[e.g., CQ Tau,][]{Ubeira2019}. 
Other transition disks also show similarly sharp density cutoffs \citep[see, e.g.][]{Brown2009,vandermarel2015}.
%{\q Brown et al. (2009) showed the inner holes in LkHα 330, SR 21N, and HD 135344B were consistent with step functions with a surface density reduction of  1000 and radii of 47, 33, and 39 AU respectively}
Planets, especially multiple planets, tend to carve only shallow cavities \citep{Duffell2015}. 

Similarly surprising is the on-going gas accretion. 
Despite inner cavities that sometimes extend to tens of AUs, the accretion rates reported for these systems are comparable to those of full disks \citep{Manara2014,vandermarel2023}. To maintain such accretion rates, gas in the evacuated regions has to lose angular momentum rapidly and falls inward with transonic or free-fall velocities \citep{Rosenfeld2014,Wang2017}. This requires efficient loss of angular momentum, and theories involving magnetized winds \citep{Wang2017} or planetary perturbations \citep{Goodman2001} have been proposed.

Motivated by these observations, we explore one physical ingredient that has so far been largely overlooked: disk shadows.

Temperature and pressure in much of a proto-planetary disk are supplied by stellar irradiation \citep{Kenyon1987,Chiang1997}. As such, any perturbations in  stellar illumination can have dynamical consequences \citep[see, e.g.][]{Watanabe2008,wu2021}. 

At the same time, azimuthal shadows 
%(cast by inner materials)
are commonly detected in  scattered light of transition disks. For example, the outer ring of J1604 shows two roughly opposed shadows \citep{Pinilla2015,Mayama2018}. These shadows are highly variable, even down to day timescales \citep{Pinilla2018}. %\x{though they remain at roughly the same  azimuths.} 
Moreover, the pair of shadow is asymmetric in their depths.
Similarly, shadow asymmetry is also reported for the TW Hya disk, one of the best monitored disks,
%and one that has been known to host one asymmetric shadow, 
where the two occulting shadows are located on the same half of the circle \citep{debes2023}.

Shadows are also reported in thermal radiation of transition disks  \citep[e.g.,][]{marel2023,Arce2023} and in chemical (C/O ratio) pattern \citep{Keyte2023,Temmink2023}.  So 
observational evidences are mounting that shadows, and in particular asymmetric shadows, are an integral part of a disk's environment. 
%These brightness asymmetries are confined to specific azimuthal angles, which {\y what is this?} can be distinguished from dust traps that are co-moving with gas. {\y are there known dust traps that comove?}

These shadows are commonly attributed to  misaligned inner disks \citep[e.g.,][]{Marino2015,Benisty2017},
%In a protoplanetary disk that hosts two mis-aligned disks, the inner disk can cast two narrow shadows or one broad shadow, depending on the relative inclination \citep{Facchini2018}. This misalignment can arise from an inclined massive planet/companion \citep{Xiang2013, Owen2017}. The precession speed therefore depends on the mass of the perturber and its orbital inclination \citep{Nealon2019}. If that is the case, the resulting shadow will move periodically instead of remaining stationary. 
%In comparison, accretion streams within the host's magnetosphere is a more promising explanation for the shadows. 3D MHD simulations show that a tilted stellar magnetic dipole can form a one-armed bending wave {\y I am not sure of this as the physical description of what happens} near the inner disk rim, where the disk scale height can be raised by $30\%$ and intercepts more light \citep{Romanova2013}. 
%{\q As matters accrete inwards, their dynamics become controlled by the increasing magnetic field so moving along the }
%In this case, the accretion takes the form of funnel flows, along two streams in the plane spanned by magnetic dipole and angular momentum vectors, so two symmetric shadows are expected on the outer ring \citep{Lyutikov2023}. This accretion rate however is not steady due to differential rotation between the inner disk edge and the star. In every few rotation periods, the inflated magnetic field will open and reconnect \citep{Bouvier2007}. 
though such a scenario may not  explain the rapid variabilities that are observed (in both magnitudes and locations). Variable accretion stream in the stellar magnetosphere is another possibility \citep{Romanova2013}.
%{\y hmm, dippers rotate by 2pi, do you want to keep it here?} 
%In addition, recent analysis of emission line radial velocity suggest a stable accretion column footprint in EX Lupi and TW Hya \citep{Sicilia2023}. 

Regardless of the origin of shadows, 
they will affect the thermodynamics and therefore the dynamics of transition disks.
%Therefore, we adopt time-independent equilibrium temperature profiles to simulate shadows in this work.
Previously, the role of symmetric shadows has been investigated by the group of \citet{Montesinos2016,Montesinos2018,Cuello2019} using 2D simulations.
In this work, we study the response of a gas ring under a time-independent, asymmetric shadow, using 3D \texttt{Athena++} simulations \citep{Stone2020}. In \S\ref{sec:method}, we describe our physical model and numerical setup. We present our results in \S \ref{sec:result}, interspersed with physical explanations and comparison against other works. We 
 discuss in \S \ref{sec:discussion} the  observational consequences.

\section{Methods}
\label{sec:method}

We perform 3D hydrodynamics simulations using \texttt{Athena++} \citep{Stone2020}.  

Many of our parameter choices are motivated by the J1604 disks, which shows a sub-mm ring at $\sim 70$ au, but shines the brightest in scattered light at $\sim 60$au. The  half width of the ring is $\sim 10$ AU \citep{Dong2017}. 
%Since J1604 is a ring peaked at 60 AU, this is a reasonable value for the scale height. 
%$ r=1.2r_0$ then $h/r = 0.1$ {\y need to make it clear in text, why choose certain parameters, correspond to J1604 obs.}
\citet{Pinilla2018} found that the two shadows, when viewed in reflected light, are roughly opposed, though their separation can vary from $150^\circ$ to $190^\circ$. The shadows also show time-varying and asymmetric depths: the eastern shadow has an amplitude that varies from 40\% to 90\%, while its  counterpart varies from 30\% to 95\%. The difference between these amplitudes can be as large as 30\%.

\begin{figure}
    \centering
    \includegraphics[width=0.9\linewidth]{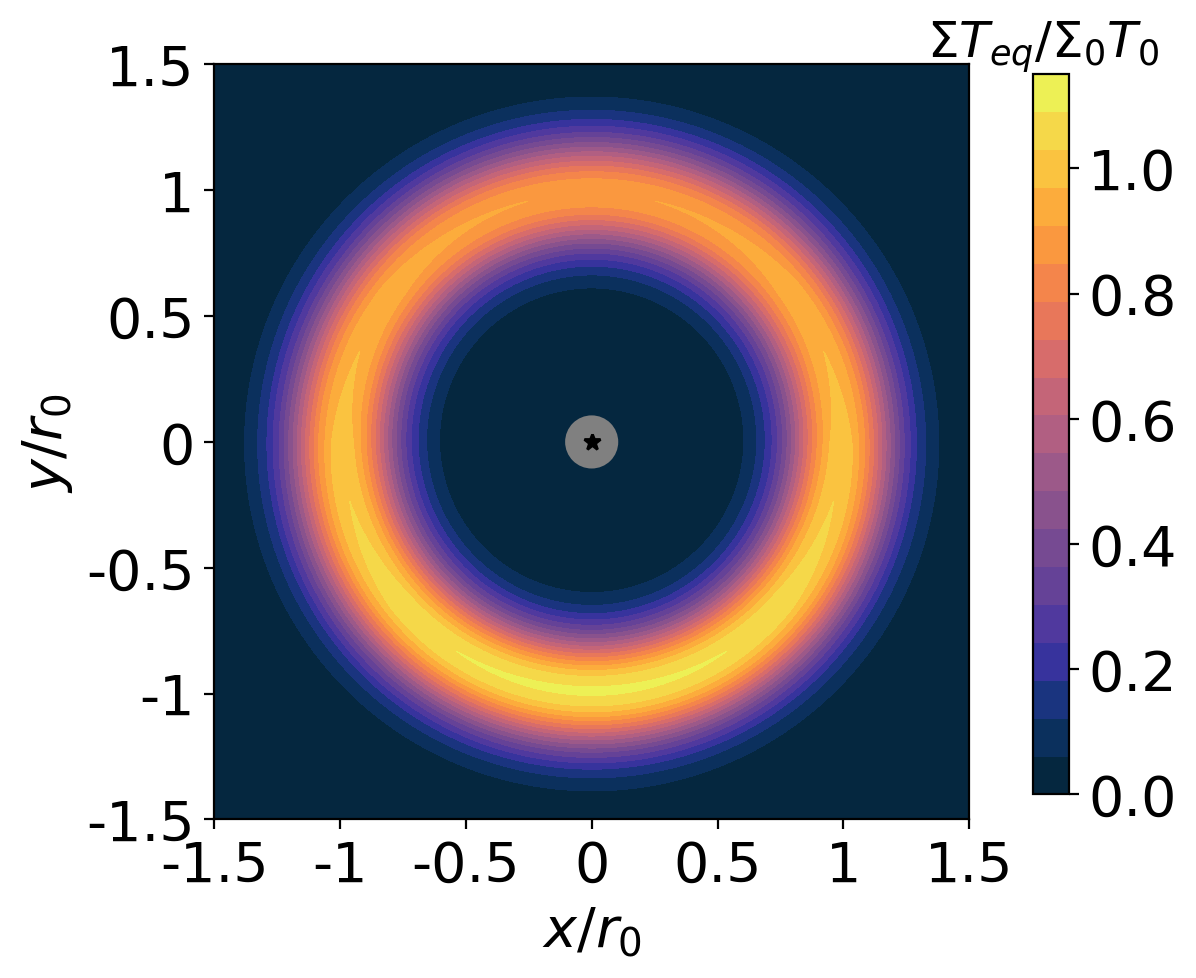}
    \caption{figure illustrating the ring and its $m=1$ shadow. The color indicates the product of the surface density (eq. \ref{eq:surfaceden}) and the imposed temperature ($T_{\rm eq}$, eq. \ref{eq:Teq}).  The latter has a $10\%$ $m=1$ perturbation on top of the axisymmetric form. The central grey circle represents the inner boundary.}
    \label{fig:enter-label}
\end{figure}

\subsection{ The Disk and Its Shadow}

We start with an axisymmetric disk. The gas surface-density is assumed to be ring-like,
\begin{equation}
    \Sigma(r) = %\frac
    {\Sigma_0}
    %{\sigma_r\, \sqrt{2\pi}}\, 
 \,     \exp\left[-\frac{(r-r_0)^2}{2\sigma_r^2}\right] \, ,
    \label{eq:surfaceden}
\end{equation}
where $\Sigma_0 = \Sigma (r=r_0)$  and $\sigma_r$ is the ring width. We choose it to be $0.16\, r_0$ in the fiducial model.\footnote{This corresponds to $\sim 10$AU if $r_0 = 60$AU. Moreover, for our chosen parameters, this corresponds to $\sim 2$ vertical scale heights at $r=r_0$.} The total disk mass is $(2\pi)^{3/2} \sigma_r \Sigma_0$. 
Computationally, simulating such a ring profile (as opposed to a full disk) has the advantage that any boundary effects from the inner and outer boundaries are minimized.

The initial vertical density distribution follows that of a vertically isothermal disk in hydrostatic equilibrium, 
\begin{equation}
    \rho(r,z) = \frac{\Sigma(r)}{\sqrt{2\pi}\, h(r)\, }\, \exp\left[-\frac{z^2}{2h(r)^2}\right]\, ,
    \label{eq:den}
\end{equation}
where the gas scale height $h (r)= c_s/\Omega$, with $\Omega = \sqrt{GM_*/r^3}$ being the Keplerian frequency and $c_s =\sqrt{k_B T/\mu m_H} $ the isothermal sound speed.
%\sqrt{k_bTr^3/\mu m_H GM}$ with the Boltzmann constant $k_b$, the mean molecular weight $\mu$ and the hydrogen atom mass $m_H$. 
The local temperature  initially scales with radius as \begin{equation}
T = T_{\rm bg} (r) = T_0 \left({r\over{r_0}}\right)^{-1/2}\, .
\end{equation}
We choose a value for $T_0$ such that $(h/r)_{r_0}=0.084$. 

The initial velocity profile follows that of pure rotation, with the azimuthal velocity satisfying the radial force balance at all heights: 
\begin{equation}
    \frac{v_\phi^2}{r} = \frac{GM_*}{r^2}+\frac{1}{\rho}\frac{dP}{d r}\, ,
    \label{eq:radial balance}
\end{equation}
where $P$ is the ideal gas pressure, $P=k_B \rho T/\mu m_H$.

Stellar irradiation is treated in a simple fashion in this initial exploration.
%{\w Our treatment of stellar illumination is similar to the one in \citet{Montesinos2016}} 
The amount of illumination is quantified by the so-called equilibrium temperature, $T_{\rm eq} (r,\phi)$. This is inserted into the energy equation of \texttt{Athena++} as an extra heating source:
\begin{equation}
    \left(\frac{dE}{dt}\right)_{\rm irradiation}
    %+\nabla \cdot \left[(E+P)\mathbf{v}\right]
    =\, \, %-v\cdot \nabla \Phi +
    - \frac{1}{\tau_c }\frac{\rho k_B \, (T-T_{eq})}{(\gamma -1)\mu m_H}\, ,
    \label{eq:extraterm}
\end{equation}
where $\gamma=5/3$ is the adiabatic index\footnote{A more appropriate value for molecular gas is $\gamma=7/5$. But this  is of no consequence here.} and we introduce the lag time, $\tau_c$, to describe the  response of the disk to a time-varying illumination.
We adopt a constant relaxation time of  $\tau_c = 0.1 P_0$, where $P_0$ is the Keplerian orbital time at $r_0$.  The disk is also assumed to remain vertically isothermal at all times.

Our two assumptions, a short thermal relaxation time, and a vertical isothermal condition, are not appropriate for the main ring of J1604. The disk's thermal inertia is high there due to the high dust density. These assumptions are more realistic inward of the main ring where the density is lower and where the scattered light is at the brightest. Here, the disk is vertically optically thin to its own thermal radiation, and radially optically thick to star light \citep[see, e.g.][]{wu2021}. This region, by definition, is also where most of the stellar heating is absorbed and so is the most dynamically relevant region.

The azimuthal shadow can now be easily accommodated by a $\phi$-dependent $T_{\rm eq}$. We assume the shadows are fixed in inertial space. We also simplify the matter by studying only one particular form of shadow. Any complicated shadows (e.g., one shadow, two symmetric ones, two asymmetric ones, broad, narrow) can be decomposed into individual  Fourier components ($e^{i m \phi}$ with $m=1,2,...$). We find that the only  Fourier component of interest is the  $m=1$ term (see \S \ref{subsec:m2} for a study of the symmetric $m=2$ term). So we write 
\begin{equation}
    T_{eq}(r,\phi) = T_{\rm bg}(r)
    %0\, \left( \frac{r}{r_0}\right)^{-1/2}\, 
    \, \left(1-\epsilon \sin\phi\right)\, .
    \label{eq:Teq}
\end{equation}
The temperature minimum and maximum lie at $\phi = \pi/2$ and $3\pi/2$, respectively.
We set $\epsilon = 0.1$. 
%For reference,  the shadows of J1604 have amplitudes in scattered lights that vary from 31\% to 95\% \citep{Pinilla2018},
%and comparable amplitudes in thermal emission \citep[880$\mu m$,][]{Mayama2012} and molecular emissions \citep[CO, HCO,][]{Mayama2018}; 
%%% I have trouble believing the ALMA data, likely due to beam dilution... 
% Zhang, Isella+2014: apparent double-lobed continuum is due to the elliptical shape of the beam, and not intrinsic asymmetry 
This is compatible with the shadow asymmetry observed in J1604 \citep{Pinilla2018}, as well as that in
TW Hya 
%range in depth from 10\% to 80\% 
\citep{debes2023}. 
%{\y is the observed scattered light or thermal? list both scattered light (Pinilla) and thermal (Mayama); thermal (Rayleigh-Jeans) is directly proportional temperature change } 
 
% The 2 shadow case consists of two Gaussians troughs at $0.5\pi$ and $1.5\pi$ respectively: 
% \begin{eqnarray}
%        T_{eq}(r,\phi) = T_0\, (r/r_0)^{-\frac{1}{2}}\, \left\{1-\epsilon\exp\left[-\frac{(\phi-0.5\pi)^2}{2\sigma_\phi^2}\right]\right\} \nonumber\\
%     \left\{1-\epsilon\exp\left[-\frac{(\phi-1.5\pi)^2}{2\sigma_\phi^2}\right]\right\} \, ,
%     \end{eqnarray}
%     {\y only $m=$ even forcing}
% with width $\sigma_\phi = 0.3$ and depth $\epsilon=0.6$. 

\subsection{Numerics}

The simulation is performed in spherical coordinates, $\{r, \theta, \phi \}$, with 
$60\times 42\times 180$ grids respectively.
In the radial direction, the domain is  $r$ $ \in [0.1\, r_0,2.5\, r_0]$, and the size of each cell increases by a factor of 1.04 in ascending order. The polar grid is evenly spaced in $\theta \in [\frac{1}{2}\pi-0.5, \frac{1}{2}\pi+0.5]$. This ensures that we cover at least 3 scale heights from the midplane for the entire disk. The azimuthal grid is evenly spaced in $\phi \in [0,2\pi]$. 

As we follow the gas up to multiple scale heights above the mid-plane, gas density in our 3D simulations has a very large dynamic range. 
%To avoid rounding errors (e.g., when converting primitive to conservative variables), {\y this reads odd}
It is common and necessary to assert floor values for density and pressure. We set them to be $10^{-9}$ and $10^{-12}$ of the mid-plane values at $r_0$. 

For the radial and polar coordinates, boundary conditions are specified in \texttt{Athena++} by the ghost cells at the  boundaries. We set $v_r=v_z=0$ at all ghost cells, to prevent gas inflow and outflow. Some attention is required at the top and bottom polar boundaries where densities are exceedingly low. Due to our adoption of a density floor, which violates the condition for hydrostatic equilibrium, waves are continuously excited in these regions. We follow \citet{Kutra2023} to mitigate this partially. We assign densities to the ghost cells as logarithmic extrapolation of the two adjacent cells. Moreover, 
the azimuthal velocities and temperatures are fixed to the initial values for these top/bottom ghost cells. For the $\phi$ direction, we adopt periodic boundary conditions.

\begin{figure}[h]
    \centering
    \vskip-0.0in
\hskip-0.15in    \includegraphics[width=\linewidth]{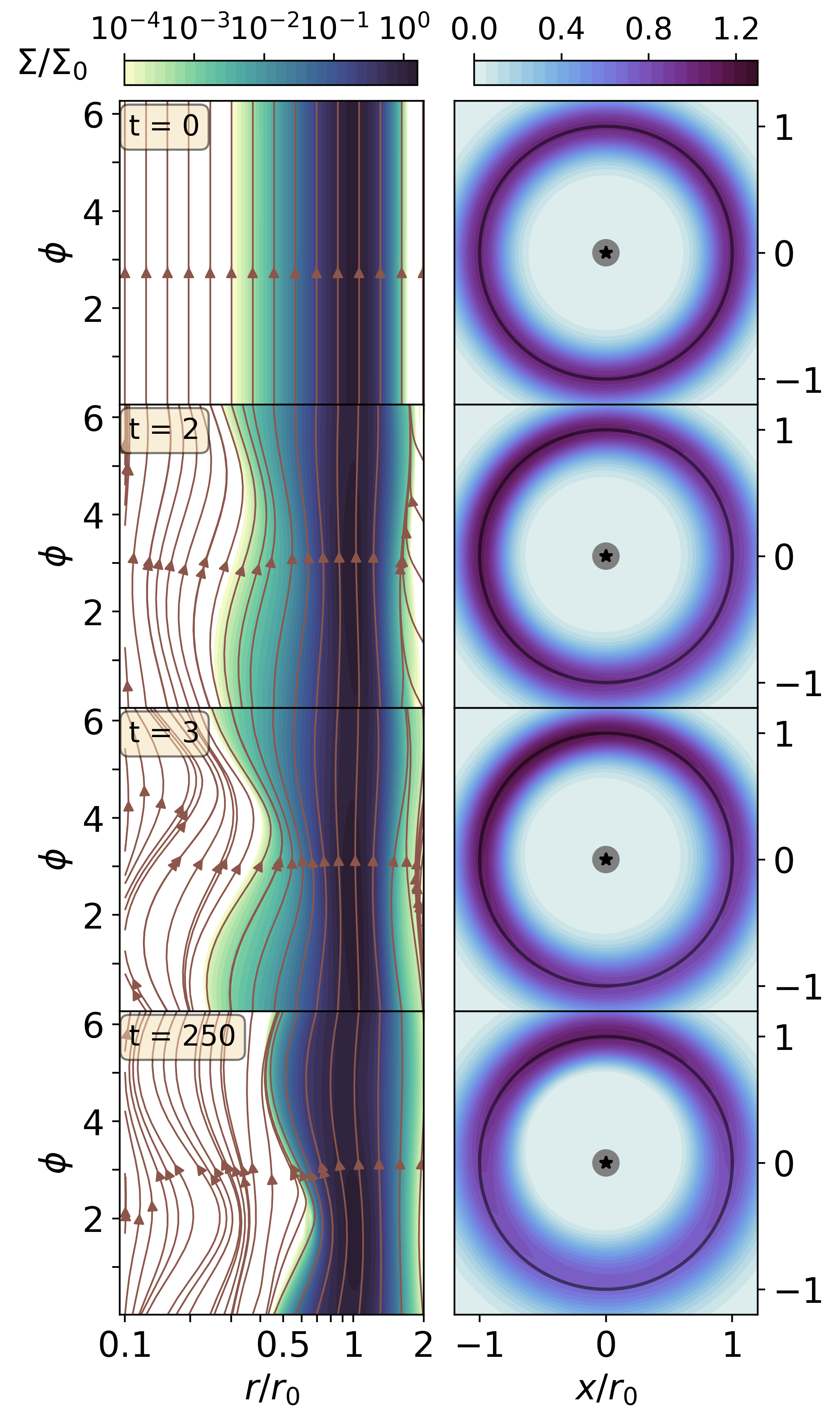}
  \caption{Four snapshots of the evolution for our fiducial ring, with time measured in unit of $P_0$. 
  %initial, linear growth, nonlinear development, final quasi-steady state. 
  On the left, the colored contours illustrate logarithmic surface densities (scaled by $\Sigma_0$) in cylindrical coordinates, while the brown curves are the fluid streamlines. 
  On the right, the colored contours illustrate the linear surface densities (again noramlized by $\Sigma_0$) in Cartesian coordinates.
  An initially circular and narrow ring reacts to the $m=1$ shadow by becoming eccentric. 
  %radially contracts near the coldest azimuth (at $\phi\sim \pi/2$ for $r\approx r_0$; higher $\phi$ values closer-in) and expands near the hottest one. 
  The eccentricity growth is complete after about a dozen $P_0$, with the final streamlines being a set of twisted eccentric ellipses. Some streamlines cross each other and truncate the disk sharply in the inner edge.
  %h of the eccentricity growth; nonlinear development; saturation. Truncation caused by streamline crossing. Final result: a sharp inner edge to the ring; eccentric (with apocentre aligned with the shadow).
}
\label{fig:narrowstream}
\end{figure}

We set a  minimum CFL number of $0.3$ and integrate the model for 250 orbital time at $r_0$ $(P_0)$. This  is equivalent to $\sim 0.1$ Myr if the ring is located at 60 AU around a solar mass star.

\begin{figure}[h]
    \centering
    \includegraphics[width=0.45\textwidth]{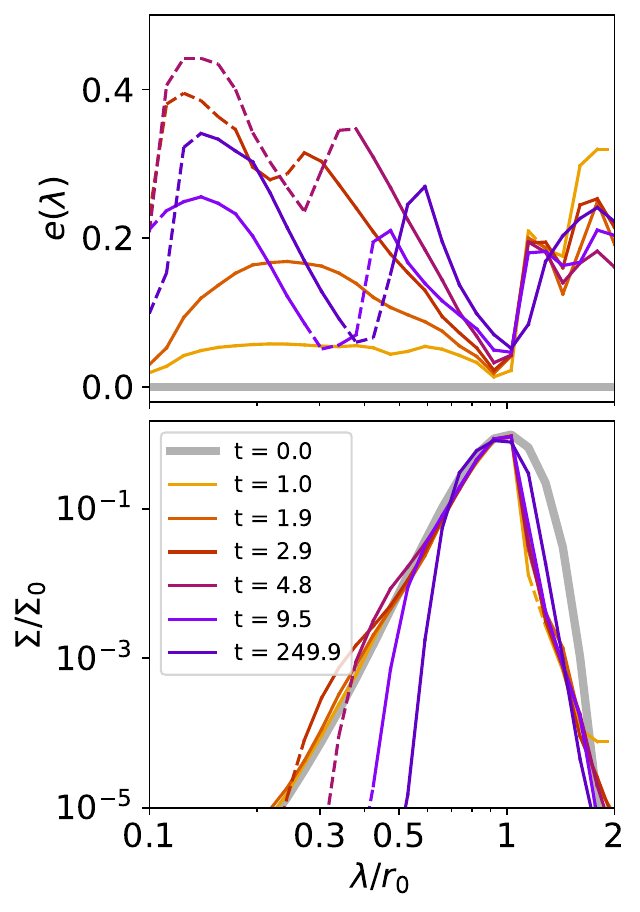}
    \caption{Evolution in eccentricity (upper panel) and surface density (lower) for the narrow, circular ring in Fig. \ref{fig:narrowstream}.
  The horizontal axis is now replaced with the semi-latus rectum $\lambda$ (eq. \ref{eq:lambda}), a more appropriate coordinate for  an eccentric disk. All results are averaged over the azimuth angle.
  %The light grey curves indicate the initial configuration, while 
  The initial profiles are plotted as thick grey curves, while the colored curves are later snapshots.
  Dashed parts indicate regions where the streamlines cross ($\partial r /\partial \lambda \leq 0$). The disk quickly becomes eccentric, and within a dozen orbits, streamline crossing has evacuated the inner region and truncated the disk sharply at $\sim 0.6 r_0$.
  }
\label{fig:firstdenecc}
\end{figure}

\section{Results}
\label{sec:result}

\begin{figure*}
    \centering
\hskip-0.1in\includegraphics[width=1\linewidth]{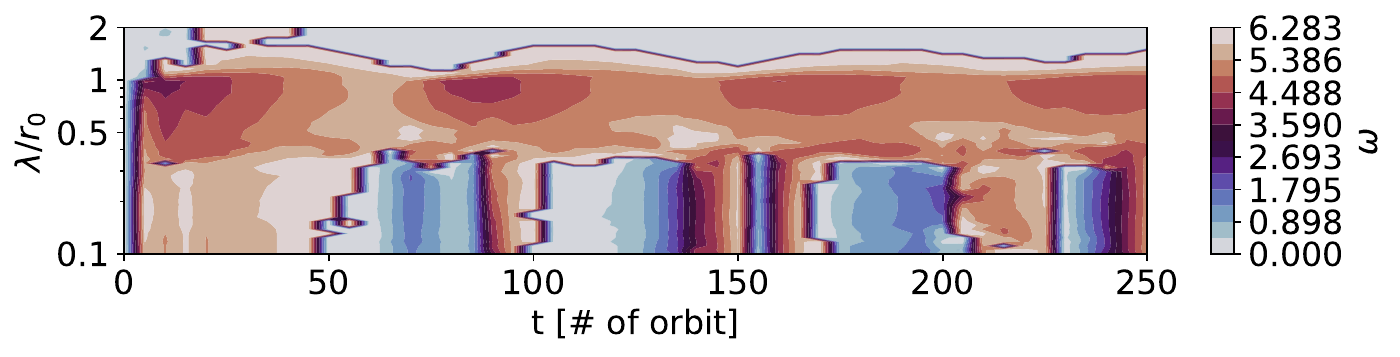}
    \caption{Evolution in the argument of periapsis (colors indicate $\omega$) throughout the disk.  The asymmetric shadow forces different streamlines to orient differently.
    Regions of the disk that develop sharp gradients in eccentricity and argument of periapsis are most susceptible for streamline crossing. The ring is quickly truncated there. During the quasi-steady state (after a dozen $P_0$), the ring continues making periodic adjustments. See text.
    %With time, 
}\label{fig:sin0.16fourier}
\end{figure*}

Here, we will describe the results of our simulations, interspersing it with discussions of the underlying physics. We find that asymmetric shadows drive the ring to become eccentric, which then leads to disk truncation and accretion. In contrast, symmetric shadows have no dynamical impacts.

\subsection{Becoming Eccentric}
\label{sec:eccentric}

Fig. \ref{fig:narrowstream} depicts how the gas streamlines evolve with time, after the asymmetric shadow is turned on. The most striking feature is that the streamlines become eccentric. 

To understand this reaction, we consider one fluid element moving in orbit. As it enters and exits the shadow, its temperature is modified, by an amount that depends on how fast it is moving and how quickly it reacts to the local forcing temperature ($T_{\rm eq}$). 
Ignoring changes in density and velocity, eq. \ref{eq:extraterm} can be re-cast as one for the fractional temperature perturbation, $\delta T/T = (T - T_{\rm bg})/T_{\rm bg}$,
\begin{equation}
{d\over{dt}}\left({{\delta T}\over T}\right) = - {1\over{\tau_c}}\left({{\delta T}\over{T}}\right) - {1\over{\tau_c}} \epsilon \sin\phi\, .
    \label{eq:dTT}
\end{equation}
We write $\phi = \Omega t$, so the forcing term (second term on the right hand side) scales as ${\cal {R}}(e^{i\Omega t - i \pi/2})$, where ${\cal R}$ indicates taking the real part. We look for a periodic solution of the form $(\delta T/T) \propto e^{i\Omega t}$. Simple manipulation yields 
\begin{equation}
{{\delta T}\over{T}} = \left[{\epsilon}\over{\sqrt{1+(\Omega \tau_c)^2}}\right] {\cal R}\left\{e^{i\Omega t+i\phi_0}\right\} \, ,
%\left[{{\Omega \tau_c - i}\over{\sqrt{1+(\Omega \tau_c)^2}}}\right]\right\}
    \label{eq:dTT_solution}
\end{equation}
where $\phi_0$ is the phase delay, $\phi_0 = {\rm arccos}[\frac{\Omega \tau_c}{(1+(\Omega\tau_c)^2)^{1/2}}]$. 
So the response magnitude is the largest 
when $\Omega \tau_c \ll 1$, or when the disk has a low density.
Moreover, with our choice of $\tau_c = 0.1 P_0$, the temperature response is nearly instantaneous at $r\geq r_0$ ($\Omega \tau_c \ll 1$), but is significantly delayed at $r\ll r_0$. This leads to elliptical streamlines that differ in their orientations (twisted).

%The gas pressure is at a maximum at the ring centre. 
The above temperature response modifies the radial pressure gradient, and causes the fluid elements to converge toward or to expand away from the ring centre, where gas pressure is at a maximum.  
In this way the $m=1$ illumination pattern excites radial epicyclic motion resonantly, driving the disk to become eccentric. In contrast, an $m=2$ pattern is not effective (see Appendix).

The magnitude of the response also depends on the background density gradient. Ignoring density changes and terms that vary slowly on scale of radius, we can simplify the radial momentum equation as 
%\begin{equation}
%{\y old \, form\, remove??}    \x{\frac{\partial \Delta v_r}{\partial t} = \Delta T\frac{\partial \ln{\rho}}{\partial r}+(\frac{\partial \Delta T}{\partial r})\, .}
%    \end{equation}
    \begin{equation}
{{\partial v_r}\over{\partial t}}%=i \Omega v_r 
\approx - {1 \over \rho} {{\partial (\rho \delta T)}\over{\partial r}} = - \delta T\,  {{d\ln \rho}\over{dr}}\, .
   \label{eq:growth}
\end{equation}
So the eccentricity ($e \sim v_r/v_{\rm kep}$) is driven the fastest %most 
where $\delta T/T$ is the highest (eq. \ref{eq:dTT_solution}), and where the background density gradient is the sharpest. 
%{\y but the problem is this gives larger $e$ at larger $r$.}
Our narrow ring with $\sigma_r \ll r$ provides the latter condition.
%Therefore, the eccentricity growth rate depends on two factors: the radial density gradient and the temperature perturbation. The former explains why the eccentricity develops at the inner edge of the disk where there is a positive density gradient. The latter suggests how the result is influenced by parameters such the shadow depth.

% \begin{equation}
%     \frac{D u_r}{D t}=-\frac{GM_*}{r^2}-\frac{1}{\rho}\frac{\partial P}{\partial r}=-\frac{GM_*}{r^2}-\frac{\partial T}{\partial r} -T(t) \frac{\partial \ln{\rho}}{\partial r}
% \end{equation}

\subsection{Nonlinear Development of the Eccentric Ring}

While the above hand-waving arguments explain why and how the disk becomes eccentric, the global and nonlinear behaviour of the disk is much less straightforward to understand. Our numerical simulations provide guidance. To analyze eccentric disks, we measure the eccentricity vector at every point by
%To study the global behavior of the disk, we firstly present the $\phi$ averaged surface density: $\int_{0}^{2\pi} \Sigma \, d\phi/2\pi$. {\z why this sentence?}
%
% The nonaxisymmetric feature can be inferred by the Fourier amplitude $C_1$
% of the $m=1$ mode in surface density.
% The coefficients are then normalized by the averaged surface density to yield:
% \begin{equation}
%     C_1 =  \frac{\int_{0}^{2\pi} \Sigma \cos{\phi} \, d\phi}{\int_{0}^{2\pi} \Sigma \, d\phi} \, .
%     \label{eq:1st moment}
% \end{equation}
%
% In addition, both the radial and tangential velocities encode information of eccentricity that are not used in Eq. \ref{eq:1st moment}. The eccentricity can be estimated by 
% \begin{equation}
%     <|\frac{v_r}{v}|>=\frac{1}{2\pi}\int_{0}^{2\pi} \left| \frac{v_r}{\sqrt{v_r^2+v_\phi^2}} \right| \, d\phi \, .
% \end{equation}
%
% \begin{equation}
%     C_{m=1} = \frac{1}{M_{disk}} \int_{r_{in}}^{r_{out}}\left| \int_{0}^{2\pi} \Sigma e^{-{\w i}\phi} \, d\phi\right| dr\, ,
% \end{equation}
% \begin{equation}
%     \varphi =  \mathrm{Arg} \left( \int_{0}^{2\pi} \Sigma e^{-i\phi} \, d\phi \right)\, .
% \end{equation}
\begin{equation}
    \mathbf{e}(r,\phi)=\left( \frac{|\mathbf{v}|^2}{\mu}-\frac{1}{|\mathbf{r}|}\right)\mathbf{r}-\frac{\mathbf{r}\cdot\mathbf{v}}{\mu}\mathbf{v} \, ,
\end{equation}
where $\mu = GM_*$. The direction of this vector points towards the periapsis, the argument of which can be  
%The eccentricity vector encodes velocity information that are not used in Eq. \ref{eq:1st moment} and t
%The angle between this and the position vector is the true anomaly, so we can obtain 
measured as 
\begin{equation}
    \omega(r,\phi) =\phi - \arccos{\frac{\mathbf{e}\cdot\mathbf{r}}{|\mathbf{e}||\mathbf{r}|}} \,.
\end{equation}
%The eccentricity vector is a constant of motion in a perfect Keplerian orbit, but in our case the temperature force causes it to vary continuously.

The $(r,\phi)$ coordinates are not convenient for eccentric orbits. We follow \citet{Ogilvie2001} to introduce a new set of coordinates, $(\lambda,\nu)$, where $\lambda$ is the semi-latus rectum $\lambda=a(1-e^2)$, and $\nu$ is the true anomaly $\nu=\phi-\omega$. 
The coordinate transformation follows the polar equation of ellipse:
%{\z $\lambda$ is used for multiple figures, but only introduced here. Either introduce it earlier, or change figures back to $r$.} {\q for the simplicity of expression, we use $\lambda$ instead of the semi-major axis.}
\begin{equation}
    r(\lambda,\phi) = \frac{\lambda}{1+e(\lambda)\cos{\nu(\lambda,\phi)}}\, .
    \label{eq:lambda}
\end{equation}
The new set $(\lambda,\nu)$ are better labels for eccentric orbits, as $\lambda$ is exactly conserved along a Keplerian orbit and approximately so for continuum fluid.
%If fluid elements are in perfect Keplerian orbits, their streamlines correspond to curves with constant $\lambda$. 
We can now proceed to plot results as functions of $\lambda$ only.

Figs. \ref{fig:narrowstream}-\ref{fig:firstdenecc} show how the eccentricity and surface density evolve in our narrow ring case. The disk reaches maximum eccentricies within a few orbits. The centre of the ring remains circular, as expected (eq. \ref{eq:growth}), while the two flanks become eccentric. 
The disk now appears as a nested set of twisted ellipses   (eq. \ref{eq:dTT_solution}). 

Such streamlines interact. One extreme form of interaction is streamline crossing.  This occurs when the derivative, $\partial r/\partial \lambda$ falls below zero, where
\begin{equation}
    \frac{\partial r}{\partial \lambda} = \frac{1}{(1+\cos \nu)^2}\left[1+\left(e-\frac{\partial e}{\partial \ln \lambda}\right)\cos \nu-e \frac{\partial w}{\partial \ln \lambda} \sin \nu\right]\, 
    \label{eq:streamline}
 \end{equation}
measures the compactness of streamlines, with $\partial r/\partial \lambda = 1$ for circular Keplerian orbits. This expression suggests that both the eccentricity gradient and the twist bring about streamline crossing. 

\begin{figure}[h]
    \centering
    \hskip-0.25in
    \includegraphics[width=1.0\linewidth]{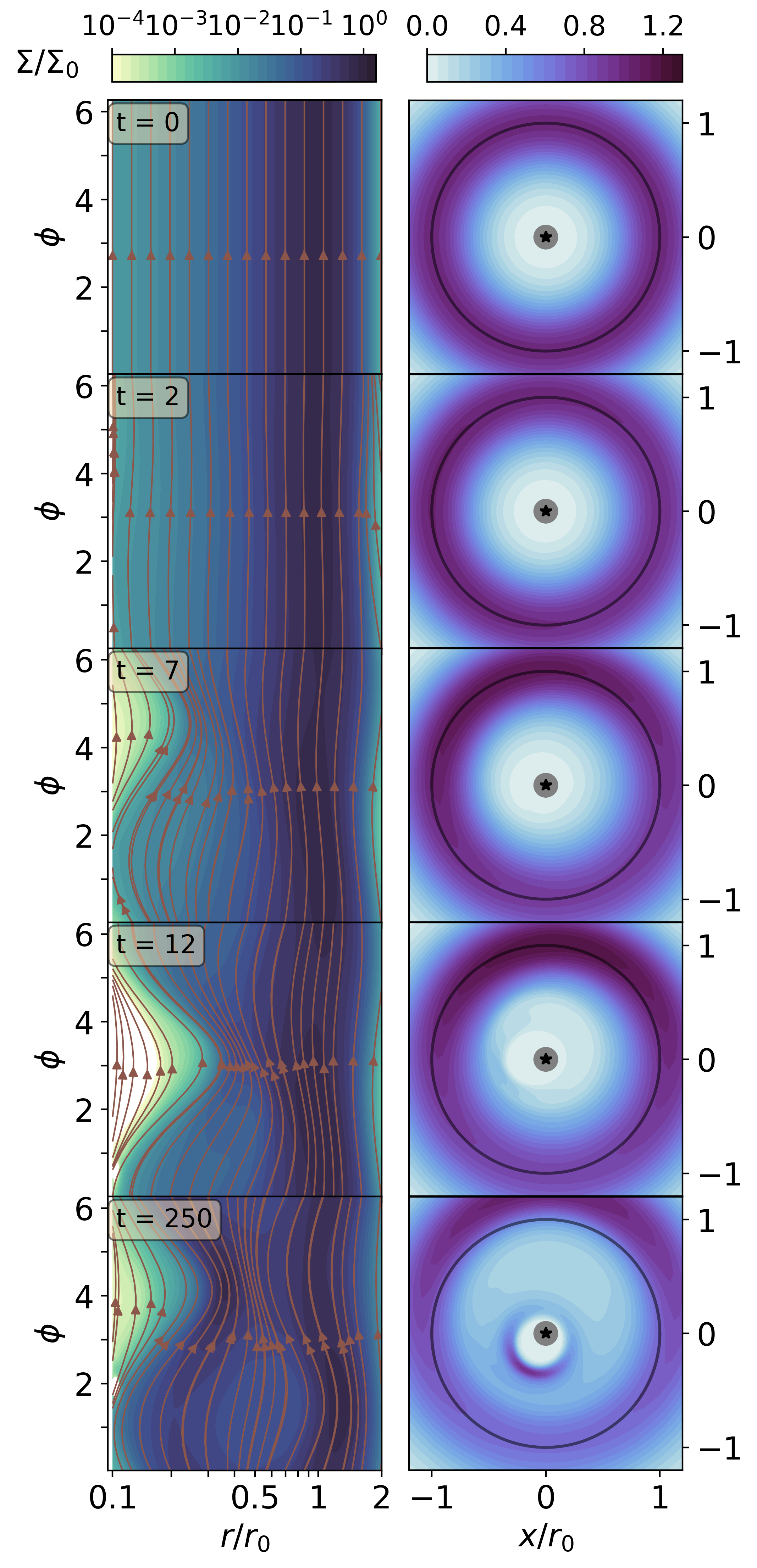}
    \caption{Similar to Fig. \ref{fig:narrowstream} but for a wider ring ($\sigma_r = 0.3r_0$). The growth of eccentricity takes longer, and a steady state is reached after about $100 P_0$. The ring is now split into two separate eccentric rings that have nearly opposite arguments of periapsis. The intermediate region is evacuated and has a density that is lower by order unity.% Surface density in the region in-between is In between, the surface density drops by $\sim 50 \%$.
     Interactions between the two eccentric rings drive continuous  mass accretion.}
\label{fig:widestream}
\end{figure}
As one observes in Fig.  \ref{fig:firstdenecc}, the low density region of the ring suffers the most from streamline crossing (marked by dashed curves). Within a dozen periods, this region is evacuated, and we are left with a ring that is sharply truncated  (more below).\footnote{The same dynamics may also be at play for the outer part. But our results there may be polluted by the proximity of the outer boundary.}

\begin{figure}
    \centering
    \includegraphics[width=0.9\linewidth]{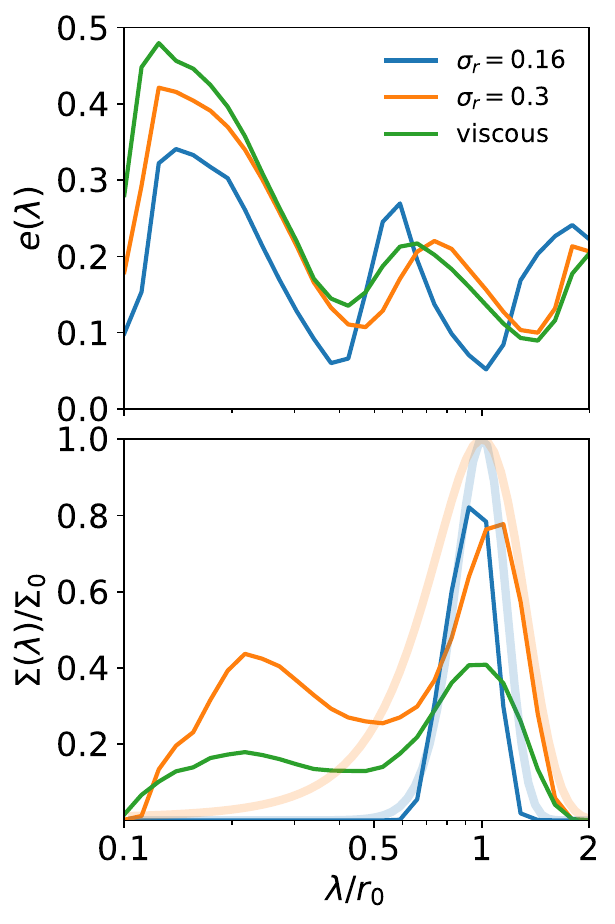}
    \caption{Final eccentricity (upper panel) and surface density (lower panel) profiles for three simulations: invisid narrow ring ($\sigma_r=0.16r_0$), invisid broad ring ($\sigma_r = 0.3 r_0$), and viscous narrow ring ($\sigma_r=0.16r_0$). 
      Initial configurations are shown as thick light curves.
    %$t=250$
All rings reach similar eccentricity profiles. Their surface densities are more different. 
Unlike the narrow invisid case which shows a sharp truncation, the wide invisid ring is broken apart into two separate rings. At the addition of viscosity, the narrow ring spreads and reaches the same state as that of the broad invisid ring.}
    %In the right panel, we find there is a critical $\sigma_r$ between 0.18 and 0.2 such that the disk exceeding this threshold will develop an inner ring.
\label{fig:deneccwidth}
\end{figure}

Afterwards, the disk settles into a quasi-steady state. The eccentricity pattern holds steady, with a minimum at $r_0$; the inner and outer halves of the ring have opposite arguments of periapsis (Fig. \ref{fig:sin0.16fourier}). This is likely the long-lived state of the ring  when under both internal stress and external forcing \citep{Ogilvie2001,Statler2001}. However, we observe that the disk is not perfectly steady. Instead, the eccentricity vectors fluctuate  with a period of $\sim 75 P_0$.  This is because the eccentric disk naturally precesses, but the forcing by the shadow does not allow it. So the disk behaves like a harmonic oscillator that is being forced away from resonance. 

\begin{figure}[h]
    \centering
\hskip0.1in\includegraphics[width=0.48\textwidth]{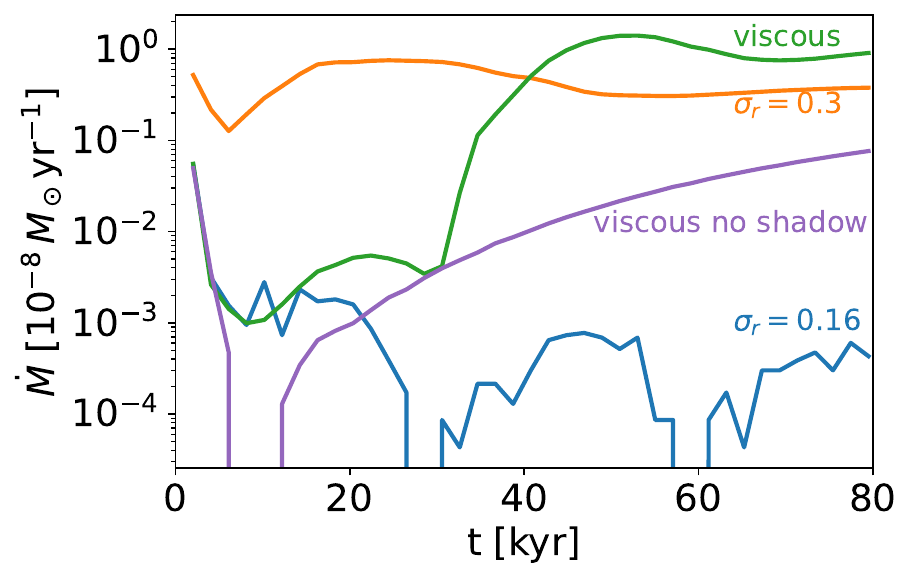}
     \caption{Mass accretion rates as functions of time in various models. Here, $\dot{M}$ is measured by the change in  total mass that remains outside $0.4 r_0$. Three of the cases are those in Fig. \ref{fig:deneccwidth}, and we add a fourth case of viscous narrow ring without shadow forcing. We assume a total disk mass of $10^{-2}\, M_{\odot}$, $r_0 = 60$ AU, $\alpha=10^{-3}$ at $r_0$, and simulate for $250P_0$. We find that asymmetric shadows alone can cause accretion (see the wide invisid case), and that in the presence of viscosity, it 
can significantly boost accretion (compare 'viscous' vs 'viscous no shadow'). 
     }
     \label{fig:massloss}
\end{figure} 

\begin{figure}
    \centering
    \includegraphics[width=0.9\linewidth]{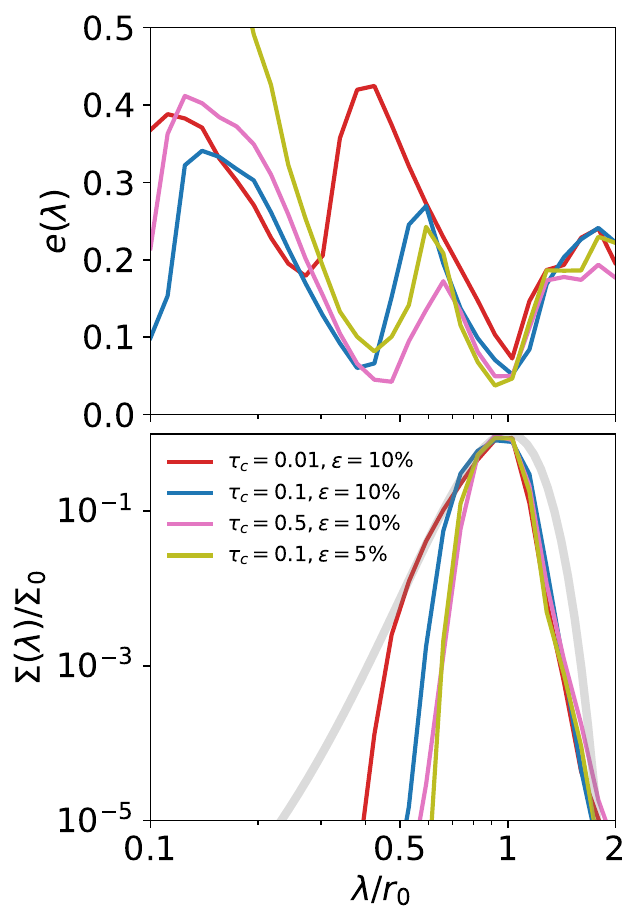}
    \caption{Final eccentricity and surface density for the narrow ring case,  now with different thermal relaxation times and driving amplitudes. The light grey curve indicates the initial state. All rings reach similar quasi-steady states as that of our fiducial case  ($\tau_c = 0.1 P_0$, $\epsilon=10\%$), namely, similar eccentricity values in the bulk of the ring, and similarly sharp truncation.  
    }
    \label{fig:parameter}
\end{figure}

\subsection{Disk Sculpting}

Here, we focus on the effects of disk sculpting. Fig. \ref{fig:firstdenecc} makes clear that the $m=1$ shadow truncates the narrow ring into one that has sharp edges. We find that the surface density drops off by more than 4 orders of magnitude just inward of $0.6r_0$, with a decaying length of order (and slightly smaller than) the local scale height. Such disks are on the verge of Rayleigh instability \citep{Papaloizou1984,Yang2010}. 
%As the disk establishes the eccentricity configuration,.... mass inside the eccentricity trough is chopped off, creating a cliff that moves outward with time. In the end, this cliff stalls at $0.4 r_0$ where the surface density is dropped by more than $10^{4}$.
So even a weak shadow is able to carve out a very sharp edge.

In this case, since our ring starts narrow, the amount of mass that is carved away (and accreted) is only a tiny fraction of the disk mass. In the following, we provide an even more dramatic demonstration of the shadow carving by simulating an initially wider ring, $\sigma_r = 0.3 r_0$. 

The results for such a  ring are shown in Fig. \ref{fig:widestream} and Fig. \ref{fig:deneccwidth}.
Eccentricity growth in a wider ring proceeds more slowly, due to the more gentle density gradient (eq. \ref{eq:growth}). We find that a quasi-steady state is reached after $\sim 100 P_0$, with an eccentricity profile that is similar to that of the narrow ring (Fig. \ref{fig:deneccwidth}).  The surface density profile, on the other hand, looks  different. We observe that the broad ring is now broken up into two separate rings with comparable surface densities. 

%As the eccentricity is built up by the shadow, neighboring streamlines are more likely to intersect at their apsides. Then the angular momentum exchange pushes materials away from the intersection and drives gas accretion. 

This occurs because the inner and the outer rings are twisted relative to each other (eq. \ref{eq:dTT_solution}). Any gas in the intermediate region suffers severe streamline crossing. 
%(with small values of $\partial r/\partial \lambda$)
%Similarly, streamlines are concentrated at the inner edge in which $\partial r/\partial \lambda$ drops below $0.2$. 
It is then whisked away and piled up instead in the inner ring. 
Fig. \ref{fig:widestream} shows that, even at steady state, the two rings continue to interact: the apoapsis of the inner ring rubs against the periapsis of the outer ring and sound waves are excited. This continuous friction leads to an interesting consequence: accretion. We  discuss this below.

%Moreover, the disk reaches a steady state without precession because these two rings ``mesh'' with each others.
 
% {\q Using $e'=-0.32$ and $w'=-2$, we estimate $e_{max}=0.37$, which is the upper limit}
% {\q this upper limit holds for simulation with different initial density.}
 
% \begin{figure}[h]
%     \centering
%     \includegraphics[width=\linewidth]{shadow1_w0.16first_eccdet.pdf}
%     \caption{Jacobian determinant of the orbital coordinates. Non-positive value indicates streamline intersection.}
%     \label{fig:firsteccdet}
% \end{figure}

In summary, an $m=1$ shadow can truncate a narrow ring sharply, and can break a wider disk into discreet rings that are eccentric and interacting.

% \begin{figure*}
%     \centering
%     \includegraphics[width=\linewidth]{shadow2_w0.16n_evolution.pdf}
%     \caption{2 shadow results. {\q Nothing interesting here {\y maybe combine Fig, 9 and 10 into one plot, showing the contrast?}}
% }
% \end{figure*}
\begin{figure*}
    \centering
    \includegraphics[width=\textwidth]{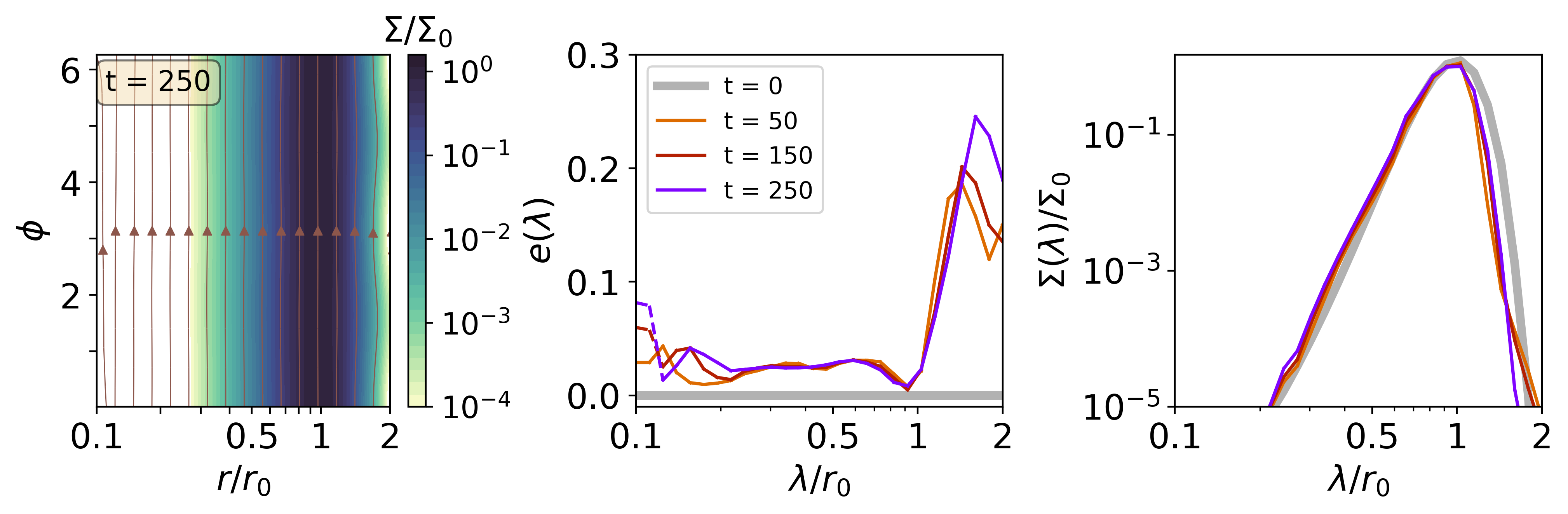}
    \caption{Evolution of the narrow ring under a symmetric shadow (eq. \ref{eq:symmetric}). The left panel shows the final streamlines. The ring remains largely circular. The middle and right panels show a few snapshots of the eccentricity and surface density profiles. Unlike for the case of an asymmetric shadow, eccentricity remains negligible and the surface density remains unchanged.}
    \label{fig:symmetric}
\end{figure*}

\subsection{Disk Accretion}

Fig. \ref{fig:massloss} presents the gas accretion rate for a few different simulations. To minimize effects of the inner boundary, we measure the mass accretion rate at $r=0.4r_0$. 

For our narrow invisid case, accretion is minimized (but non-zero) after the ring has reached the quasi-steady state. However, a healthy rate of accretion is maintained in the wide invisid case, despite the absence of viscosity.
%Without viscosity, the narrow ring experiences a negligible amount of accretion. 
The continuous 'rubbing' between the twisted inner and outer rings causes repetitive compression and decompression of the streamlines, much like Sisyphus and his impossible task of rock lifting. This transports angular momentum and, as we find here, leads to accretion. We find that the accretion rate is comparable to that of a standard $\alpha$-disk (see below).

Even when the disk is viscous, shadow forcing dramatically enhances the accretion. 
We demonstrate this by simulating the narrow ring with viscosity, and compare the cases with and without shadow forcing. We adopt a constant kinematic viscosity of $\nu=10^{-5}$ (in code unit), which corresponds to a conventional $\alpha$ parameter of $\alpha \sim 10^{-3}$ at $r_0$ and $\sim 10^{-2}$ at $0.1 r_0$ \citep{Shakura1973}.
% {\q given constant $\nu$, $\alpha \propto 1/r$?} 
The shadowed ring accretes at a rate that is about five times higher. The accretion is likely enhanced by the sharp density gradient that an asymmetric shadow produces.

%Similar to the wide rings, the mass is continuously transferred to the center by two processes: the viscosity broadens the ring, and the streamline compression efficiently removes mass at the inner edge.

\subsection{Parameter Dependence}

Our results depend on two important parameters: the thermal relaxation
time, $\tau_c$, and the amplitude for shadow forcing, $\epsilon$. 

In Fig. \ref{fig:parameter}, we present the final eccentricity and surface density profiles when the values of $\tau_c$ and $\epsilon$ are modified from our fiducial ones. 
To our surprise, we find that the final states are not sensitive to these parameters. 

We have expected these parameters to affect the magnitudes of temperature perturbations (eq. \ref{eq:dTT_solution}), producing weaker responses when  $\tau_c$ is longer, and when $\epsilon$ is smaller. And we do observe that in  simulations, rings take longer to build up their eccentricities if $\tau_c$ are larger and if  $\epsilon$ are smaller. However, when these rings reach their quasi-steady states, their eccentricity profiles for the main ring (near $r_0$) look largely invariant. All  disks are also truncated with a similar sharpness. 

We suspect this convergence is the result of nonlinear evolution. The shadow forcing saturates at the same final configuration, regardless of the amplitude of forcing.  These results should be followed up with more in-depth investigations. Meanwhile, it gives us some confidence that our results are not sensitive to the choice of the parameters. 
%The cooling timescale $\tau_c$ can also impact the growth because it averages out the temperature perturbation along the orbit. To investigate the dependency, we preform addition two simulations with $\tau_c=0.01 P_0$ and $\tau_c = 0.5 P_0$. As predicted by Eq. \ref{eq:dTT}, the disk takes longer to build up the eccentricity with larger $\tau_c$. 
%At the same time, the final eccentricity at the inner edge is smaller. {\q truncation radius?}

%{\w when amplitudes reduced to $5\%$, eccentricity lower, inner ring less mass, accretion rate lower; so for observed transition disk at 50au, $10\%$ amplitudes required for dramatic evolution}

% \begin{figure*}[h]
%     \centering
%     \includegraphics[width=\linewidth]{shadow1_nu-5n_surfacexymulti.pdf}
    
%      \caption{Surface density of a viscous ring.
%      Mass is transferred to the center at faster rate.
%      }
%      \label{fig:sin0.16nu}
% \end{figure*}     

\subsection{ Symmetric Shadows: comparison with other Studies}
\label{subsec:m2}

Here we first show that symmetric shadows do not drive eccentricity. As symmetric shadows are the default set-up in a number of previous studies and a variety of dynamical effects have been reported for them, we discuss the discrepancy.

For the symmetric shadow, we assume it to consist of two Gaussians troughs at $0.5\pi$ and $1.5\pi$ respectively: 
\begin{eqnarray}
       T_{eq}(r,\phi) & =&  T_0\, \left({r\over{r_0}}\right)^{-\frac{1}{2}}\,\times 
   \left\{1-\epsilon\exp\left[-\frac{(\phi-0.5\pi)^2}{2\sigma_\phi^2}\right]\right\}\nonumber \\
   & & \times
    \left\{1-\epsilon\exp\left[-\frac{(\phi-1.5\pi)^2}{2\sigma_\phi^2}\right]\right\} \, .
    \label{eq:symmetric}
    \end{eqnarray}
In terms of Fourier components, this shadow can be decomposed into $m=2, 4, 6, 8...$ components.  We choose a shadow-width 
$\sigma_\phi = 0.3$ and a relatively large shadow-depth of $\epsilon=0.6$. The ring is radially narrow with $\sigma_r =0.16$. Fig. \ref{fig:symmetric} shows the evolution (or rather, the lack of) under this shadow. The ring stays largely circular and the surface density profile remains unchanged. The eccentric mode is suppressed because the forcing imposed by symmetric shadows is not resonant with the orbit. 

These results contrast with those reported by previous studies, all of which focus only  on symmetric shadows and employs 2D simulations (vertically integrated equations).

\citet{Montesinos2016} simulated the effects of a pair of symmetric shadows on a continuous disk instead of the ring. Their very deep shadows (blocking out $0.999$ of the starlight) can excite shallow density spirals, when the disk is gravitationally unstable, or when the stellar luminosity is unphysically high. In contrast, there is little effect when the disk is low in mass or when the stellar luminosity is modest. This is consistent with our results here. 

Later studies from the same group invoke additional effects like slowly precessing symmetric shadows \citep{Montesinos2018}, or back-reaction from dust particles on the shadowed disk \citep{Cuello2019}. These appear to produce interesting effects but are not in direct contradiction with our results.

%\citet{Montesinos2018} simulate 2 shadows moving in prograde direction. Planet-like features are found at the corotation radius. (transition disk)
%\citet{Cuello2019} find also trace dust particles and find pressure difference in spiral can trap dust. (transition disk)

Lastly, a preprint appeared at the same time as our work \citep{Su2024}. These authors simulated disks with a wide range of physical parameters. They found that symmetric shadows on continuous disks (not rings) can trigger structures like spirals, rings and crescents. It is hard to understand the difference between our conclusions and theirs, before one develops a clear physical explanation for the reported effects.
In the meantime, we speculate that both nonlinearity (deep as opposed to the shallow shadows studied here) and boundary effects (continuous disks experience the inner boundary) may play some roles.

%Method difference: cooling (beta cooling and \citet{Montesinos2015} cooling by assuming blackbody) and dynamical range of negative power law)

%The simulation of wider ring in the right column of Fig. \ref{fig:sin0.16fourier} shows a different evolution pattern: a significant amount of mass is transferred to the inner ring. After 30 orbits, the surface density breaks at $r=0.5$ and $e$ shows a local minimum. 
    
\section{ Connections to observations}
\label{sec:discussion}

Our exploratory study reveals the surprising benefits of an asymmetric shadow. In the following, we speculate on possible connections to the observed systems.

\subsection{Disk Sculpting}

We return to the issue of sharp inner edges. 
Disk shadows, even with an asymmetry of order a few percent, can sharply truncate a ring within tens of orbits. In our narrow ring case, the density at the edge falls off by many orders of magnitude, with a characteristic scale that is the local scale height. This can be maintained for as long as the shadows are present. This could help explain systems like J1604 and others (see Intro) without the need to invoke massive companions.

In addition to sharp truncation, our simulations also show that a wide ring can be dismantled into two pieces with comparable surface densities. It is also possible, if we are able to simulate a larger radial domain, that an initially smooth disk can be shattered into  multiple rings. Such multiple ring system are commonly observed  in high resolution images, with famous examples like HL Tau \citep{alma2015}, AS 209 \citep{Huang2018} and TW Hya \citep{Andrews2016}. Whether shadow is the main cause of these sub-structures require more investigations.

%{\w abundant information on sharp rings cutting off with local scale heights, from ALMA}

%{\w formation of inner ring}

\subsection{Eccentric disk}

There are at least two known eccentric disks. Both MWC 758 and AB Aur are confirmed to be eccentric and off-centered in mm wavelengths \citep{Dong2018, marel2021}. MWC 758, in particular, has a measured eccentricity of 0.1, similar to what we find in simulations (Fig. \ref{fig:deneccwidth}). A few more candidates of eccentric transition disks are recently revealed by \citet{jenson2023} from spectro-astrometric observations of CO ro-vibrational lines.
%\footnote{However, it is also possible that the line emission is affected by shadows.}
%There are also examples of brightness asymmetries in mm wavelength, which could be a sign of apocenter glow \citep{pan2016}, their eccentricity is too small and generally unconstrained due to the resolution. 

If eccentric transition disks are wide-spread, asymmetric shadows provide a universal mechanism that is natural and efficient. In this case, 
we predict that disk eccentricities tend to be greater further away from the ring centre, and the arguments of periapsis are not aligned. These could be tested by observations of dust and gas. In this regard, the recent report of an asymmetric ring around CIDA 9A \citep{Harsono2024} gives one the hope that high-resolution, multi-wavelength observations would be soon available to test the shadow hypothesis.

Lastly, the dynamics of grains in such eccentric disks is worth investigating.  For grains that are not strongly coupled to the gas, they may not follow the gas streamlines closely. There can be complicated gas-dust interactions (strong wind, fragmentation, drifting) in eccentric disks.

%However, small grains are elevated from the midplane so factors such as inclination and scale height cause degeneracy in eccentricity measurements.

%The dynamics of dust grains in such an eccentric disk are unclear. The crashing of gas streamlines may lead to particle fragmentation. This complex gas motion may also impact the radial drifting rate of grains, potentially concentrating solids at particular locations. Consequently, it is difficult to draw conclusions about what to expect from dust observations.

%{\q dust observation possibly due to eccentric gas dynamics \citep{Harsono2023}(arxiv) }

\subsection{Accretion in Transition Disks}

If transition disk objects do accrete at healthy rates, material in the central cavities would have to lose angular momentum rapidly and move inwards at appreciable speeds. Interacting streamlines, forced by asymmetric shadows, can either act as the agent of friction itself (as in our wide ring inviscid case), or as the facilitator for viscosity (as in our narrow viscous case). 

%Observations show that both  gas and dust are heavily depleted in the cavities of transition disks \citep{Calvet_2005 ,marel2016}, so the high accretion rate of transition disk requires gas to flow at transsonic speed inside the cavity. This require strong gravitational torques, exerted from undetected giant planets. 

% The high accretion rate of transition disk and other types of protoplanetary disk requires the transport of gas angular momentum. Source of turbulence is MHD and VSI. Our results rise a new possibility because streamline crossing can remove angular momentum
% In addition, the shadow can amplify the accretion with small viscosity. 

Shadow forcing has another advantage over alternative scenario like magnetized wind \citep{Wang2017} or planetary perturbations \citep{Goodman2001}. It can work at all radii where shadows are cast, i.e., over many decades in orbital separations. 

Streamline crossing occur likely at supersonic speed. If this is present in disk cavities, one can hope to find signatures in grain sublimation and gas emissions.

%Our results offers a new way to remove the gas by the crushing of eccentric streamlines. In addition, the shadow can amplify the accretion in the disk with the low level of turbulence constrained by observation \citep{Flaherty2017,Flaherty2018}. 

\subsection{Kinematic Signatures}

%AA Tau       ALMA found 59deg outer rings; w/ HCO+ in inner region appearing to be rotated relative to the outer disk (https://arxiv.org/abs/1704.02006), like J1604

%Juhasz+Fachini’17 modelled with a warped inner disk; 

%Rosenfeld+ modelled with a radial inflow that is free-fall in HD 142527

\begin{figure*}[ht]
%     \begin{subfigure}{8cm}
%    \centering
\includegraphics[width=8.5cm,trim={0.1cm 0.2cm 0 0.cm},clip]{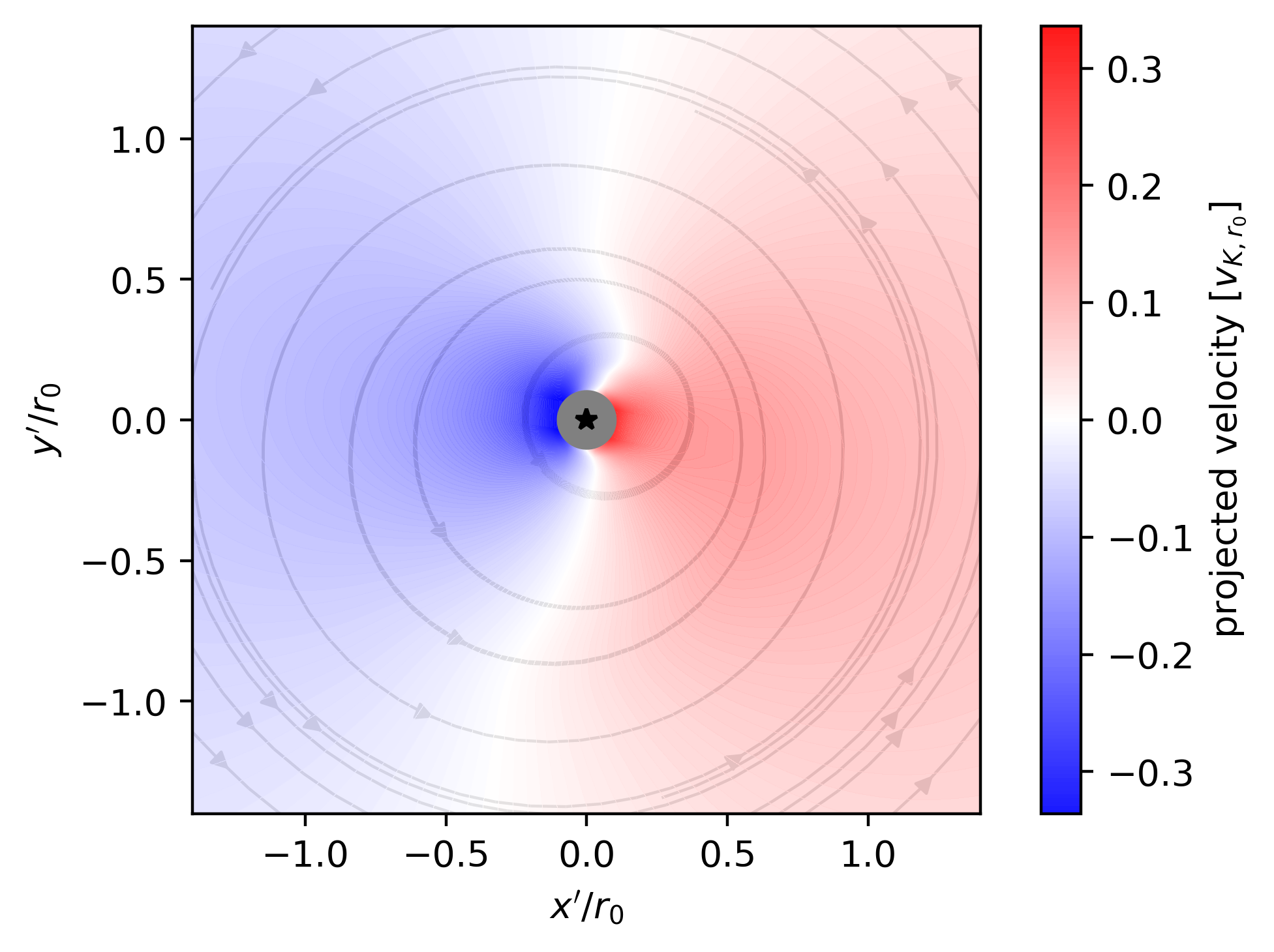}
%    \caption{}
%     \end{subfigure}
%    \begin{subfigure}{8cm}
%    \centering
\hskip0.5in\includegraphics[width=7cm,trim={0.1cm -0.64cm 0 0.25cm},clip]{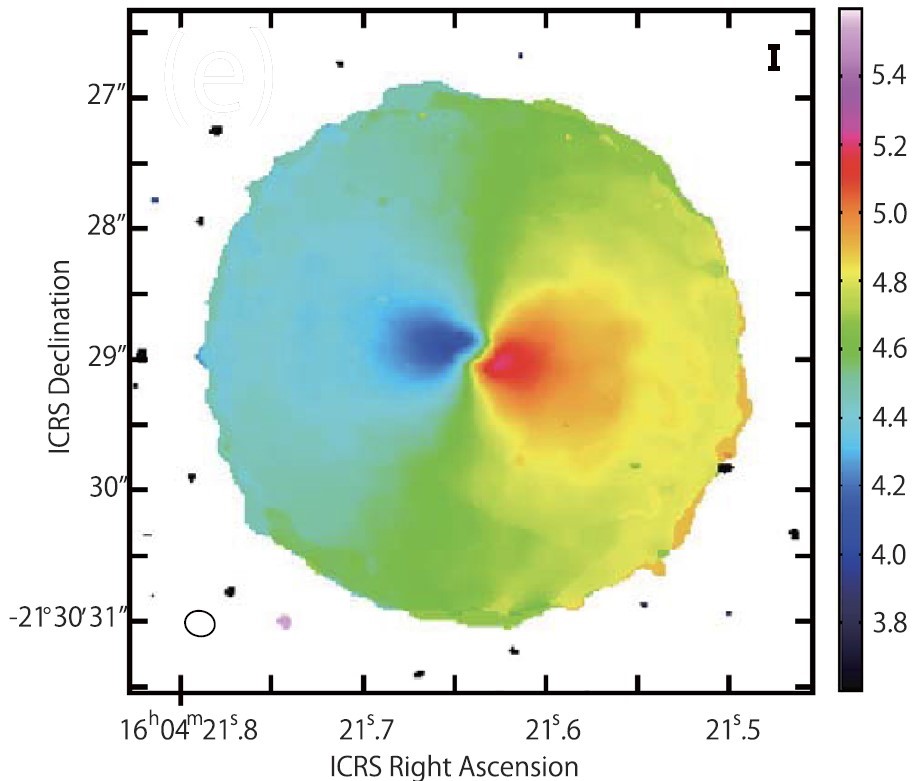}
%    \caption{CO (3-2) moment 1 map of RXJ1604 from Fig. 1e in \citet{Mayama2018}}
%    \label{fig:1604co}
%     \end{subfigure}
     \caption{Kinematic signature of a twisted eccentric disk. The left panel shows the projected velocity of our wide ring case ($\sigma_r = 0.3r_0$, $r_0 = 1$, at $t=250P_0$), with the grey contours indicating the velocity streamlines. %our integration boundary. The contour marks the surface density level of $\Sigma/\Sigma_0=[0.4,0.65]$. 
     The velocity is in unit of circular Keplerian speed at $r_0$ ($\sim 3.8\km/\s$ at 60au). The right panel is the  moment-1 map of J1604, reproduced from Fig. 1e of \citet{Mayama2018} with permission. The unit on the color bar is km/s. The two are visually comparable.}    
     \label{fig:projvel}
    \end{figure*} 

The eccentric streamlines should also be observable in kinematically resolved line emissions. 

In the centre of many transition disks, line emissions are found to be distorted in the so-called
intensity-weighted velocity maps (moment-1 maps). These include, e.g. HD 100546 \citep{Walsh2017}, DoAr 44 \citep{Antilen2023}, an SY Cha \citep{Orihara2023}. Fig. \ref{fig:projvel} presents the case of J1604,  where both the CO (3-2) and $HCO^+$ (4-3) moment-1 maps show a twisted structure inside the cavity \citep{Mayama2018}.

This feature has been interpreted as due to warped inner disks \citep{Loomis2017, Juhasz2017}, or due to fast radial inflow \citep{Rosenfeld2014}. 
We suggest here that  it could also result from eccentric streamlines in an inclined disk. 

We produce a theoretical image for visual comparison.
We adopt the final snapshot of the wide ring case ($\sigma_r=0.3$). We then calculate the projected line-of-sight velocity:
\begin{equation}
    v(x',y') = v_\phi \sin{i}\sin{\phi}-v_r\sin{i}\cos{\phi}\, ,
\end{equation}
in the frame $(x',y')=(x\cos{i},y)$ where $i$ is the disk inclination and $\phi$ the azimuth angle. The left panel of Fig. \ref{fig:projvel} shows the projected midplane velocity when $i = 6^\circ$, the estimated value for J1604's disk \citep{Pinilla2018inc}. 
The velocity map is not symmetric with respect to the disk minor axis. Instead the blue-shifted emission occupies more region than the red-shifted counterpart. In addition, the zero velocity line deviates from the minor axis and is twisted in a similar way as the observed CO (3-2) moment-1 map of J1604 \citep{Mayama2018}.

% \begin{figure}[h]
%     \centering
%     \includegraphics[width=\linewidth]{shadow1_w0.25_projvel.pdf}
%     \caption{Projected line-of-sight velocity of 1 shadow model with $\sigma_r=0.2$ at t=150. {\w suggest to use j1604 viewing geometry, and perhaps put Mayama figure next to this?}}
%     \label{fig:projvel}
% \end{figure}

%{\q limitation: CO and $HCO^+$ are optically thick, so choosing midplane is not accurate and vertical velocity also contributes.}

\section{Summary \& Outlook}

In this work, we explore how uneven stellar illuminations influence  transition disks. 
Our 3D hydrodynamic simulations show that asymmetric shadows, even measured a few percents in depth, drive the disk to become eccentric. This in turns leads to streamline crossing, disk truncation and accretion. We show that this may explain a number of observed oddities of transition disks, including inner edges that are as sharp as allowed by physics, eccentric rings, rapid accretion across the cavity, and twisted velocity maps. In particular, a sharp edge can be carved without the presence of a massive perturber, and a healthy accretion can be driven even in the absence of disk viscosity.

These surprising benefits of shadows are not known before, but they represent perhaps only a small facet of the interesting dynamics in irradiated disks. Many more in-depth studies are needed, even for our particular set-up of a shadowed transition disk.  For instance, our crude assumption of a constant lag time should be abandoned. Instead, the disk response to changes in the stellar illumination is more nuanced, with dependencies on factors such as surface density, vertical location, disk flaring, etc. We have also assumed that disk shadows are asymmetric and fixed in inertial space. While supported by available observations, unless we have a first-principle theory for the origin of shadows, these are at best guesses at their characters. Lastly, we need global simulations that span across a larger dynamic range in radius.
%The inner disks found in wide rings are subjected to the inner boundary effect which can be improved with more dynamical range. If the inner disk is robust, shadows can continue to truncate it and form multiple rings. If it is not real, materials flow into the star with highly inclined orbits, leading to significant mass accretion.

Shadows can wreck havocs across a large swath of radial distances, without the need to invoke massive perturbers or disk viscosity. This makes them, in our view, a promising and potent new driver in proto-planetary disks. Targeted observations should be performed to test this theory. The trademark of shadow, as we discover in this work, is the twisted eccentric streamlines.
%Such process is more efficient on narrow disk where the disk evolves to a eccentric ring with sharp inner edge within a few orbital time. It takes the wide disk around 50 orbital time to evolve into two region in which the inner region possess a different argument of periapsis. A preliminary derivation shows the growth of eccentricity and $\omega$ gradient facilitate the crossing of streamlines. Once streamlines pass over each others, mass is pushed away from the intersection and the angular 

%\section*{Acknowledgements} 
\bigskip
The authors thank an anonymous referee for many useful comments, and thank Chris Thompson, Norm Murray, Chris Matzner, Janosz Dewberry and Xuening Bai for helpful conversations. They also acknowledge funding from NSERC. 

\bibliography{sample63}{}
\bibliographystyle{aasjournal}
\end{document}